\documentclass{pasj00}

\begin{document}
\SetRunningHead{Tamura et al.}{The Millimeter Sky Transparency Imager (MiSTI)}
\Received{2010/11/25}
\Accepted{2010/12/27}
\Published{}

\title{The Millimeter Sky Transparency Imager (MiSTI)}

\author{%
	Yoichi \textsc{Tamura}\altaffilmark{1,2}, 
	Ryohei \textsc{Kawabe}\altaffilmark{1,3},
	Kotaro \textsc{Kohno}\altaffilmark{4,5}, 
	Masayuki \textsc{Fukuhara}\altaffilmark{2}, 
	Munetake \textsc{Momose}\altaffilmark{6}, 
	Hajime \textsc{Ezawa}\altaffilmark{7,1}, 
	Akihito \textsc{Kuboi}\altaffilmark{2,8},
	Tomohiko \textsc{Sekiguchi}\altaffilmark{7,9},
	Takeshi \textsc{Kamazaki}\altaffilmark{10,7}, 
	Baltasar \textsc{Vila-Vilar\'{o}}\altaffilmark{10},
	Yuki \textsc{Nakagawa}\altaffilmark{6}, 
	and
	Norio \textsc{Okada}\altaffilmark{7}
}
\altaffiltext{1}{Nobeyama Radio Observatory, National Astronomical Observatory, Nobeyama, Minamimaki, Minamisaku, Nagano 384--1305}
\email{yoichi.tamura@nao.ac.jp}
\altaffiltext{2}{Department of Astronomy, University of Tokyo, Hongo, Bunkyo, Tokyo 113--0033}
\altaffiltext{3}{Department of Astronomical Science, The Graduate University for Advanced Studies, 2--21--1 Osawa, Mitaka, Tokyo 181--8588}
\altaffiltext{4}{Institute of Astronomy, University of Tokyo, 2--21--1 Osawa, Mitaka, Tokyo 181--0015}
\altaffiltext{5}{Research Center for the Early Universe, School of Science, University of Tokyo, Hongo, Bunkyo, Tokyo 113--0033}
\altaffiltext{6}{College of Science, Ibaraki University, Bunkyo 2--1--1, Mito, Ibaraki 310--8512}
\altaffiltext{7}{National Astronomical Observatory, 2--21--1 Osawa, Mitaka, Tokyo 181--8588}
\altaffiltext{8}{Kasukabe Girls' High School, 6--1--1 Kasukabe Higashi, Kasukabe, Saitama 344--8521}
\altaffiltext{9}{Asahikawa Campus, Hokkaido University of Education, Asahikawa, Hokkaido 070--8621}
\altaffiltext{10}{Joint ALMA Office, Av.\ Alonso de Cordova 3107, Vitacura 7630355, Santiago, Chile}

\maketitle


\begin{abstract}
The {\it Millimeter Sky Transparency Imager (MiSTI)} is a small millimeter-wave scanning telescope with a 25-cm diameter dish operating at 183~GHz. MiSTI is installed at Atacama, Chile, and it measures emission from atmospheric water vapor and its fluctuations to estimate atmospheric absorption in the millimeter to submillimeter. 
MiSTI observes the water vapor distribution at a spatial resolution of $0.5 \arcdeg$, and it is sensitive enough to detect an excess path length of $\lesssim 0.05$~mm for an integration time of 1~s.
By comparing the MiSTI measurements with those by a 220~GHz tipper, we validate that the 183~GHz measurements of MiSTI are correct, down to the level of any residual systematic errors in the 220~GHz measurements. 
Since 2008, MiSTI has provided real-time (every 1~hr) monitoring of the all-sky opacity distribution and atmospheric transmission curves in the (sub)millimeter through the internet, allowing to know the (sub)millimeter sky conditions at Atacama. 
\end{abstract}


\section{Introduction}
\label{sect:intro}

The major obstacle that leaves the millimeter (mm) to sub-millimeter (submm) windows still unexploited in ground-based 
astronomy is severe atmospheric absorption mainly caused by tropospheric water vapor. 
Moreover, small scale non-uniformity in its spatial distribution distorts wavefronts of astronomical signal, 
limiting the natural seeing at these wavelengths (e.g., \cite{Carilli99}).
At the ALMA \citep{Wootten09} site, precipitable water vapor (PWV) contents of the atmosphere have been the subject of several site testing campaigns using devices such as  tipping radiometers, Fourier-transform spectrometers (FTSs) and water vapor monitors (WVMs) operating at mm to submm wavelengths. These experiments succeeded in modeling atmospheric transparency at these wavelengths as a function of precipitable water vapor \citep{Matsuo98, Matsushita99, Paine00, Paine04, Pardo01}. 

Another important fact from the viewpoint of telescope operation at mm and submm wavelengths 
is that observations above $\sim 450$ GHz usually require very good weather conditions that are even at an excellent site like the Chajnantor plain in Chile. Dynamic scheduling is, therefore, necessary to make best use of a telescope under all conditions, and real-time monitoring of the atmospheric condition is indispensable  to execute it.  Tippers operating at 220--225~GHz are widely used at submm observatories (e.g., CSO, ASTE, APEX, etc.) for sounding the sky opacity and assessing astronomical data quality. 
An all-sky mid-infrared (IR) imager  to monitor clouds on the sky, similar to those commonly used at optical/near-IR observatories
\citep{Takato03, Shamir05, Suganuma07, Sebag08, Miyata08},  is also working at the ASTE site (4860~m in elevation; Pampa la Bola, Atacama, Chile; \cite{Ezawa04, Ezawa08} for the ASTE facility). 
There has been, however, no instrument to map the mm/submm emission of water vapor across the sky with high sensitivity. 

It will be important to monitor not only the amount of absorption by water vapor but also the amplitude of small-scale ($\sim$10--100~m) fluctuations because they govern the astronomical seeing at mm and submm wavelengths (see  \citet{Altenhoff87} for single dish observations and \citet{Thompson01} for interferometry). The turbulent layer of water vapor is known to be $\approx$1~km above the ground level of the Atacama plateau \citep{Delgado01, Robson02, Beaupuits05}.

\begin{figure*}
\begin{center}
	\FigureFile(61mm,1mm){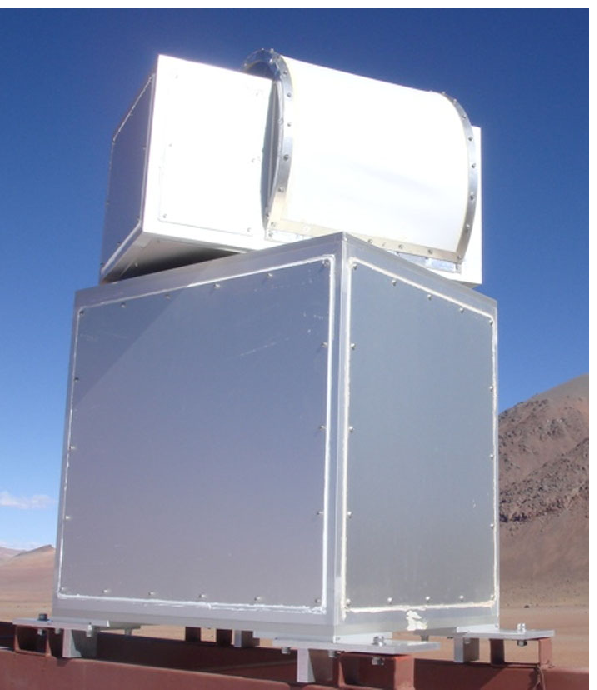}\hspace{7mm}%
	\FigureFile(55mm,1mm){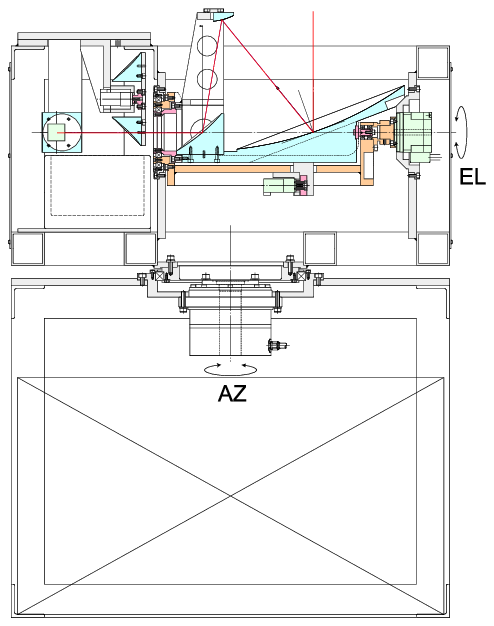}
	\end{center}
		\caption{
			(\textit{left}) A picture of the Millimeter Sky Transparent Imager (MiSTI) installed
			at the ASTE site. The elevation drive unit equipped with the primary, secondary, 
			and tertiary mirrors is covered with a thin GORE-TEX membrane to avoid desert 
			dust and snow. 
			(\textit{right}) Schematic drawing of MiSTI. The membrane and its support structures 
			are not shown. The components in cyan are optical mirrors, and the line in red 
			shows the optical axis. The positions of the azimuth and elevation axes are noted 
			by ``AZ'' and ``EL'', respectively.
		}\label{oshima}
\end{figure*}

\begin{figure*}
	\begin{center}
		\FigureFile(42mm,41mm){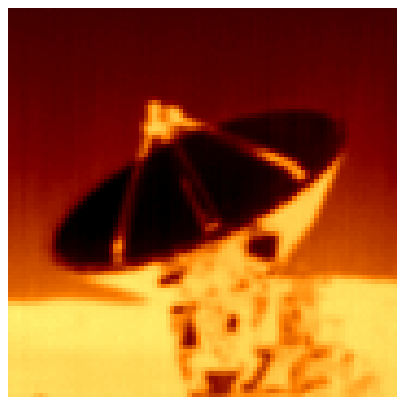}%
		\FigureFile(42mm,41mm){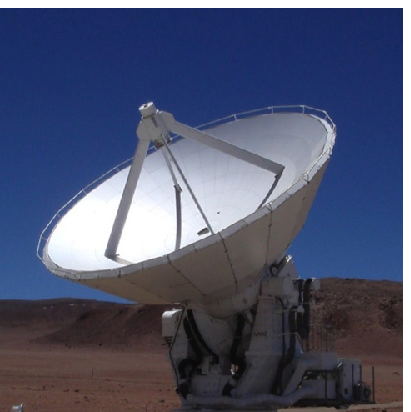}
	\end{center}
		\caption{
			(\textit{left}) The first-light image of MiSTI.  The target object is the ASTE 10-m dish. 
			The receiver of MiSTI was tuned at a local-oscillator frequency of 196~GHz, where 
			the atmospheric transmission is higher than at 183~GHz. The brighter colors represent 
			higher intensities at 196~GHz.  We also show the optical image of ASTE (\textit{right}). 
		}\label{FirstLight}
\end{figure*}

The {\it Millimeter Sky Transparency Imager (MiSTI)} is a 25-cm diameter telescope operating at 183~GHz installed at the ASTE site. It is designed to measure the emission from the atmospheric water vapor and its time variation in any arbitrary azimuth ($Az$) and elevation ($El$) direction. A picture and schematic drawing of MiSTI are shown in figure~\ref{oshima}. 
and monitoring, and using it to verify the quality of ALMA data (i.e., data quality assurance). 
We have developed MiSTI at the National Astronomical Observatory of Japan (NAOJ), and have installed and tested the instrument at the high-altitude ASTE site. 
The `first light' at the ASTE site was received in December 2007 (figure~\ref{FirstLight}), and stand-alone automatic/remote operations were started in March 2008. Since May 2008,  a dedicated web site for MiSTI\footnote{\texttt{http://aste-www.mtk.nao.ac.jp/\symbol{"7E}misti/opacity.html}} has been opened to provide the real-time 183~GHz sky images to radio astronomers who already have (and will have) telescopes in Atacama. 

In this paper, we present the instrumental overview of MiSTI and its initial results; more detailed analyses of the sky measurements will be presented in a subsequent paper. Requirements on the instrument and methods for measuring water vapor are introduced in  Section~2.  The system overview of MiSTI is described in Section~3. Application of the MiSTI as an all-sky opacity monitor is presented in Section~4. Sections~5 and 6 devote discussions and summary, respectively.

\section{Requirements and Methodology}
\subsection{Science Goals and Requirements}

One of the scientific goals of MiSTI is to map and monitor the mm emission of water vapor across the sky to use it for dynamic scheduling of astronomical observations and data quality assurance. This allows an observer at submm observatories to choose a transparent region of the sky to be observed and to make the observing schedule more efficient. This requires high sensitivity to water vapor and the ability to rapidly point to arbitrary positions of the sky, since all-sky mapping with a single beam is generally time-consuming. 

The other goal is to understand small-scale ($\sim$10--100~m) structure of water vapor distribution $\approx$1~km above the ground level. This can only be done with a radiometer by quickly measuring the amount of water vapor at $\approx$2--9 points within a small patch of the sky, a 1$\arcdeg$--5$\arcdeg$ square region for instance. The measurement is on an experimental basis and will be very challenging. But it is important because the small-scale fluctuation of water vapor is believed to affect natural seeing of submm single-dish telescopes and interferometers. This requires a spatial resolution of $\sim 0.5\arcdeg$ and a quick scanning or wobbling ability. Furthermore, sensitivity to water vapor and a high stability receiver are necessary to measure the tiny fluctuations of water vapor even for short integration times of $\sim$1~s. Portability of the instrument is also required since experiments may be performed at different observatories and sites.

The required sensitivity to water vapor and diffraction-limited spatial resolution depend on observing frequency. In the next section we examine the optimal observing frequency.


\subsection{Selection of Observing Frequency}
\label{sect:waterVapor}

One of the most effective ways to detect small variations of water vapor is to observe a water line. In the wavebands which have a significant water line, water vapor emission increases the closer to the line centers. Furthermore, the amount of variation in brightness temperature for a unit variation in the amount of water vapor also increases, which improves the sensitivity to tiny changes in PWV. Water vapor emits several lines at centimeter to submm wavelengths (e.g., 22, 183, 325~GHz). We choose the 183~GHz line for the following reasons. The 22~GHz line is always optically thin in the dry conditions at Atacama, and hence the sensitivity to variations of PWV is lower than at 183~GHz, as suggested by \citet{Wiedner01}. Moreover, receivers at 325~GHz generally have higher noise temperatures than 183~GHz. 

In order to optimize the radio frequency that is efficient to detect small variations of water vapor, we use an atmospheric transmission model \textsc{atm} \citep{Pardo01} to characterize the sky brightness temperature and excess path length in the 183~GHz band. 
We assume an altitude and vertical pressure distribution similar to those in Atacama. PWV is set to 0.8, 1.4, and 2.8~mm, which are the first, second and third quartiles (i.e., 25\%, 50\% and 75\%) of PWV in Atacama, respectively \citep{Radford98}. 

\begin{figure*}
\begin{center}
	\FigureFile(80mm,1mm){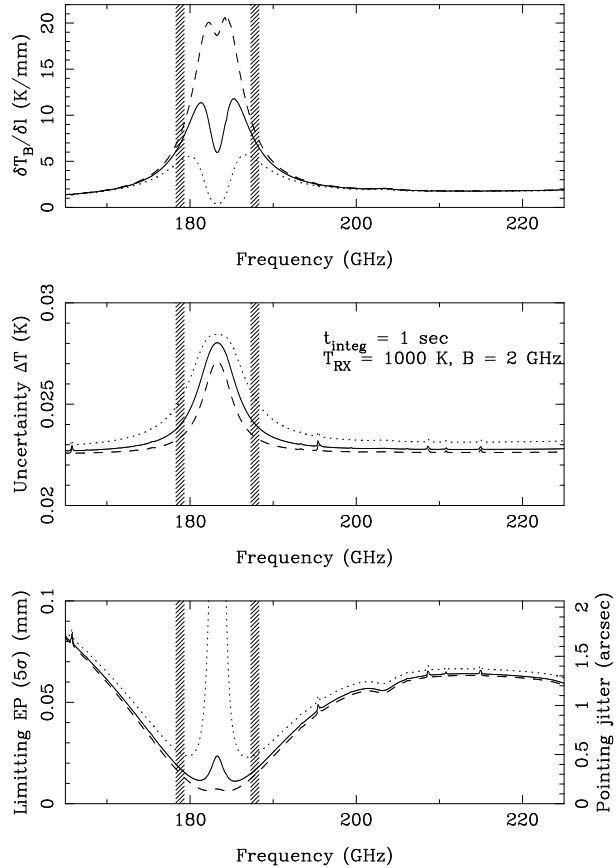}
	\end{center}
		\caption{
			(\textit{top}) The change in brightness temperature of the sky per unit change in excess
			path (EP), $\delta T/\delta l$. The dashed, solid, and dotted curves show $\delta T/\delta l$
			for the first, second and third quartiles (i.e., 25\%, 50\% (median), and 75\%) of PWV 
			at Chajnantor, respectively. The hatched regions indicate the frequencies at which 
			MiSTI observes the sky. 
			(\textit{middle}) The uncertainty of sky brightness measurement as a function of 
			frequency when assuming the integration time $t_\mathrm{integ} = 1.0$~sec, the 
			receiver noise temperature $T_\mathrm{RX} = 1000$~K and the double-sideband 
			bandwidth $B = 2$~GHz. 
			(\textit{bottom}) The effective sensitivity (5$\sigma$) to EP length $l$ as a function of 
			frequency. The scale of the corresponding detection limit of wavefront tilt (pointing 
			jitter for a 10-m telescope) is indicated in the right axis. 
			See Section~\ref{sect:waterVapor} for details.
		}\label{frequency}
\end{figure*}

 In the top panel of figure~\ref{frequency}, we show the variation in brightness temperature per unit change in excess path length, $\delta T/\delta l$, as a function of frequency. The middle panel shows the uncertainties in sky temperature measurement, $\Delta T$, as a function of frequency. The uncertainty is expressed as $\Delta T = (T_\mathrm{RX} + T_\mathrm{b}^\mathrm{sky}) / \sqrt{B t_\mathrm{integ}}$, where $T_\mathrm{b}^\mathrm{sky}$ is the sky brightness temperature, and we assume a receiver noise temperature $T_\mathrm{RX} = 1000$~K, the band width $B = 2$~GHz (DSB) and the integration time  $t_\mathrm{integ} = 1$~s. 
 If we consider the former and latter as the signal and noise, respectively, then effective sensitivity to water vapor emission is proportional to a ratio between $\delta T/\delta l$ and $\Delta T$. The bottom panel of figure~\ref{frequency} shows the minimum detection limit to excess path length (EP). The local minima of the detection limit appear at $\nu _\mathrm{opt} \approx 183.3 \pm$(3--4)~GHz, and the position of the local maxima slightly moves outward from the water line center with increasing PWV. This is because water vapor emission at frequencies close to the line center is rapidly saturated with increasing PWV and therefore it becomes relatively insensitive to variation of PWV. 
We expect that the water line is almost saturated for PWV $>$ 2.8~mm at 180--186~GHz. This situation is more likely to occur when the telescope points at low elevation angles. We prefer the frequency band in which $\delta T/\delta l$ is sufficiently large and the water vapor emission is not saturated over a broad range of PWV. We therefore adopt $183.3 \pm 4.5$~GHz as the observing frequency of MiSTI. 

We find that a 183~GHz receiver operating at room temperatures with $T_\mathrm{RX} \approx 1000$~K and $B = 2$~GHz is sensitive enough to detect an EP of $\lesssim 0.05$~mm for 1~s integration. Note that if a difference in EP at two different points of the sky separated by a distance of 10~m is 0.05~mm, a wavefront falling on a 10~m telescope aperture can tilt at $\sim 1''$ (see the right axis of the bottom panel in figure~\ref{frequency}).

\section{The System}
\label{sect:design}

\subsection{Telescope}

\begin{figure*}
	\begin{center}
		\FigureFile(110mm,1mm){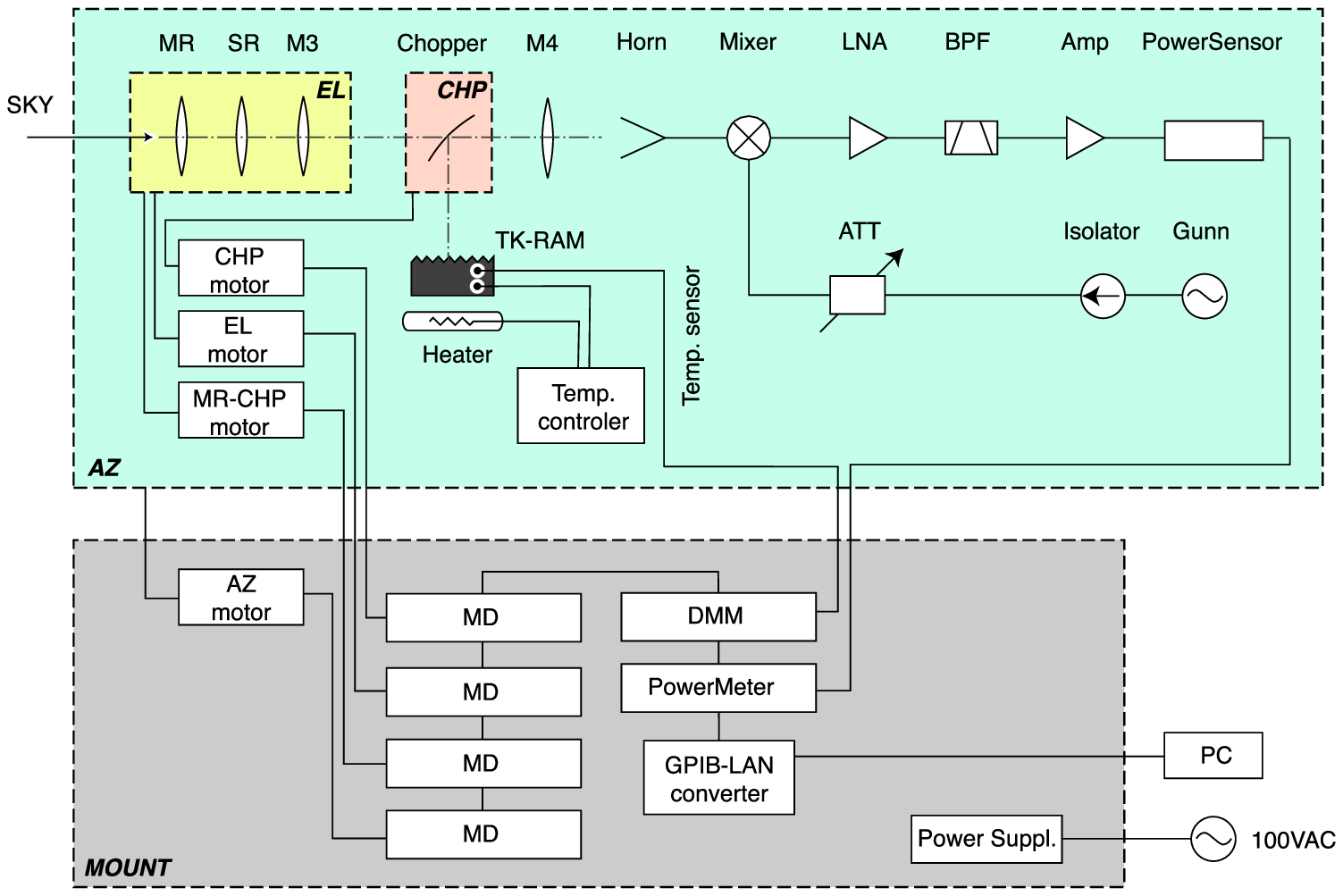}\hspace{5mm}%
		\FigureFile(45mm,1mm){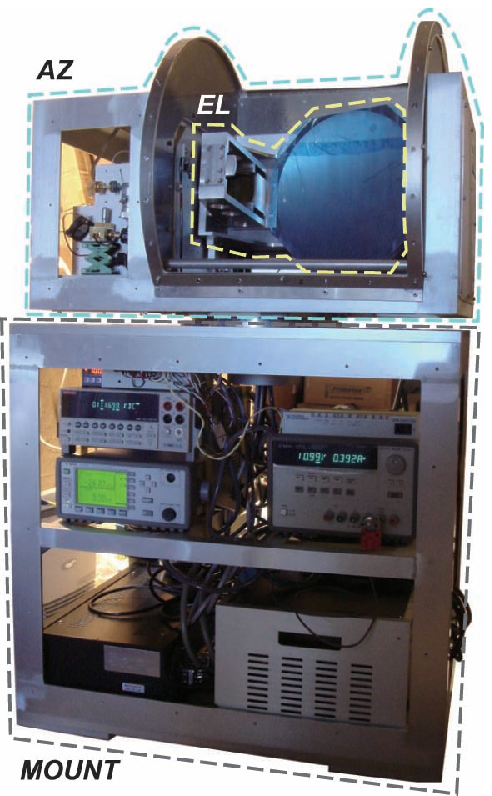}
	\end{center}
		\caption{
			(\textit{left}) The system block diagram of MiSTI. The boxes outlined by dashed 
			lines indicate the mount and instrument room (grey box), the azimuth unit (pale-green),
			the elevation unit (yellow) and the chopping mirrors (orange),  and they are independently 
			driven by servo motors and their drivers (MDs). 
			Emission from the sky is collected by main reflector (MR), sub-reflector (SR) and the third 
			and forth mirrors (M3, M4). Then, the emission is coupled by a feed horn, and detected 
			by the receiver and the power sensor. The reference signal at 183.3~GHz is generated 
			from a Gunn oscillator, and coupled at the mixer. 
			For flux calibration, the chopping mirrors (CHP) bend the beam fed by the receiver, 
			and couples the beam with loads (TK-RAM), whose temperatures are controlled and/or 
			monitored by the temperature controller and digital multimeter (DMM). 
			The power meter, the digital multimeter, and the motor drive units are controlled from 
			a Linux machine via a local area network at the ASTE site. 
			(\textit{right}) A picture of the inside of MiSTI. The dashed lines in grey, pale-green, and 
			yellow indicate the mount, the azimuth unit, and the elevation unit, respectively. 
			The primary mirror in the picture is temporarily covered with a blue protective film.
		}\label{system}
\end{figure*}

MiSTI is a 1-m tall, portable mm-wave telescope with a 25~cm primary mirror. The specification and system block diagram are shown in table~\ref{tab:specification} and figure~\ref{system}, respectively. The optics and receiver of MiSTI is on an alt-azimuth mount. The telescope boresight can move in the ranges of $Az=-175\arcdeg$ to $+175\arcdeg$ and $El = 0\arcdeg$ to $95\arcdeg$. For the measurements of small-scale structure of water vapor, the elevation drive unit equipped with a main reflector, secondary and tertiary mirrors can be wobbled at an angle of 30$'$ at 1~Hz.  This allows to measure PWV at two points on the sky, $\approx$9~m apart from one another if the water vapor layer is 1~km above the ground. 
The elevation drive unit is outside the receiver cabin and it is covered with a 0.5-mm-thick GORE-TEX membrane to avoid desert dust and snow (see figure~\ref{oshima}). The instruments for measurement and control, except for a control computer,  are inside the lower mount. 

MiSTI is installed on a frame firmly welded on the top of a container at the ASTE site. The latitude, longitude, and altitude are N22$\arcdeg$58$'$18$''$, W67$\arcdeg$42$'$10$''$, and 4860~m, respectively. The infrastructure (e.g., power supply and internet access) is provided by the ASTE facility.

\subsection{Optics}
The optics of MiSTI is a dual-offset modified-Nasmyth design, and consists of a main paraboloidal reflector (MR), a secondary hyperboloidal reflector (SR), two ellipsoidal mirrors (M3 and M4) and a feed horn. The main reflector is an 250-mm offset paraboloidal mirror, and it has a focal length of $f = 245.0$~mm. The main reflector is under-illuminated: the edge taper level is $T_\mathrm{e} = -20$~dB, which provides an angular resolution of 30~arcmin at 183.3~GHz and low ($\lesssim$1\%) spillover loss. The beam shape was checked by raster-scanning the Sun. The sky emission collected by the primary, secondary and tertiary mirrors is guided into the receiver cabin along the elevation axis, and it is fed into the mixer through the corrugated feed horn after being focused by the forth mirror. For antenna temperature calibration, MiSTI has two offset-paraboloidal mirrors, each of which moves and chops the ray at a beam waist between M3 and M4. Then, each chopping mirror bends the beam from the feed to couple it with radio absorbers at different temperatures. All of the optical elements are optimized to 183.3~GHz on the basis of Gaussian optics.

\subsection{Front-end and back-end system}

The front-end system consists of a diode mixer, a low-noise amplifier (LNA), a bandpass filter (BPF), and a second amplifier, all of which are operated at room temperature (i.e., not cooled). We use a subharmonic mixer, WR-4.3SHM (Virginia Diode, Inc.), which has a DSB noise temperature and a conversion loss of 600~K and 5.5~dB, respectively, over the operating frequencies 175--220~GHz. The local oscillator (LO) signal is supplied from a Gunn oscillator (WiseWave Technologies, Inc.). The Gunn oscillator has no phase-lock loop, since we found that the LO frequency always falls within a narrow frequency range ($\lesssim$100~MHz) centered at 183.3~GHz. Furthermore, we made sure that the uncertainty in opacity estimates is very small ($<1\%$) for any conditions at PWV = 0.2--2.0~mm even if the LO frequency shifts by 100~MHz from the water line center. This is because a shift of LO frequency can increase power received by one of the sidebands and decrease power received by the other sideband, which almost compensates the total detected power. An attenuator inserted between the oscillator and mixer is used to optimize the power of the LO signal. The intermediate frequency (IF) is centered at 4.5~GHz with a bandwidth of 1.0~GHz. 
The receiver noise temperature is $T_\mathrm{RX} \approx 900$~K. 
The biases of the amplifiers and oscillator are fed by a direct-current (DC) power supply (E3631/Agilent Technologies, Inc). 

We employ a power sensor and meter (E9300A and E4418B/Agilent Technologies, Inc.) as an IF total power detector and its readout instrument. The readout signal detected with a power sensor is modulated (i.e., AC-coupled) to separate the signal from the low-frequency noise. This significantly improves suppression of $1/f$-type noise. 

\begin{figure*}
	\begin{center}
		\FigureFile(75mm,1mm){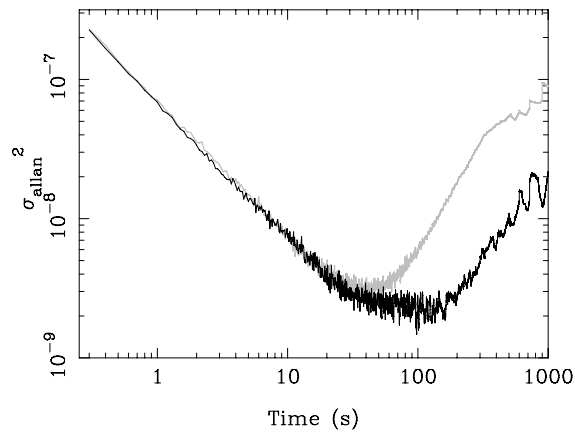}
	\end{center}
	\caption{
		The Allan variance of receiver output at 196~GHz. The feed was coupled with a radio 
		absorber at a temperature of 297~K. We measured the data with turning on only devices 
		for receiver operation and data acquisition (i.e., power meter, DC supply, and network 
		devices; black curve) and all devices for full operation (i.e., power meter, DC supply, 
		network devices, motor drivers, hot load, and fan; grey curve).
	}\label{allanplot}
\end{figure*}

Figure~\ref{allanplot} shows the Allan variance of the output of the power meter. We measured the output at 196~GHz for 3600~s only turning on the devices used for receiver operation and data acquisition. The feed was coupled with a radio absorber at a constant temperature of 297~K. The power meter data were normalized by the power averaged over the measurement. The Allan variance shows a thermal behavior down to $\sim$100~s where it starts to flatten at a level of $\sigma_\mathrm{allan}^2 \sim 2 \times 10^{-9}$. This indicates a gain stability of $\Delta G/G \approx \sqrt{2 \times 10^{-9}} \sim 5 \times 10^{-5}$ for a time duration of $\lesssim 100$~s. When the devices for full operation (including motor drivers, hot load system, and fan) are powered on, the gain stability slightly degrades probably due to electromagnetic interference from those devices. We, however, find $\sigma_\mathrm{allan}^2 \sim 3 \times 10^{-9}$ for 50~s, suggesting that the gain is still quite stable ($\Delta G/G \lesssim 10^{-4}$) for 50~s.

\subsection{Intensity Calibration}
\label{sect:calibration}
For measurements of antenna temperature of water vapor emission, the dual-load calibration method is appropriate for intensity calibration since we do not need prior information about atmospheric temperature, which is often difficult to know. MiSTI is equipped with dual-temperature loads in its receiver cabin. One is operating at an ambient temperature, and the other is heated. Temperatures of both of the loads are monitored. The beam radiated from the feed is folded by rotating paraboloidal mirrors, which focus the beam on temperature standards. By making the radiated area on the loads as small as possible (beam radius on the loads, $w = 3.6$~mm), the load temperature measurements are easier and more accurate. 

We use Radio Absorbing Material (RAM; Thomas Keating Inc.) for temperature standards. We insert a high-precision temperature sensor module (LM35CA; National Semiconductor) in each of the RAMs to monitor their temperatures. The absolute accuracy is $0.4\arcdeg$C over an operating temperature range from $-40$ to +110$\arcdeg$C. 

We have paid special attention to keeping the temperature of the hot load as stable as possible.
The hot load system consists of a RAM, a flat heater (SAMICON~230; Sakaguchi E.H Voc Co.), a temperature control module (TTM-000; Toho Electronics Inc.) and a small circuit for safety and temperature control. The radio absorbers are settled on an aluminum holder, which has high thermal conductivity and is used for heating the RAM from its back and sides to make the surface temperature of the absorber uniform. We filled the gap between the absorber and the holder with silicone grease to improve the thermal conductivity. The window to the RAM is covered with a foam polystyrene sheet (1~mm in depth) to keep the surface temperature of the RAM high and constant. The temperature controller monitors the RAM temperature with a thermocouple, and it keeps the temperature constant with a proportional-integral-derivative (PID) control algorithm. We tested the hot load system and found that the system can stably and safely heat up the absorber up to $\approx$110$\arcdeg$C. 

An optional cold temperature standard can be available. The receiver cabin has a space in which a shallow dewar flask with RAM at the bottom is set. The flask can be filled with 1~litter of liquid nitrogen to cool the RAM. The chopper mirror can bend the beam and illuminate the cooled RAM. The boiling point of liquid nitrogen at 4860~m altitude (air pressure of 570--575~hPa) is $\approx 73$~K.

For measurements of atmospheric transparency by sampling powers from the sky at several or more elevation angles, commonly referred to skydip measurements, the one-load calibration method is still useful. In the one-load method, an atmospheric temperature is assumed to be equal to the ambient load temperature. The formulation is given in Section~\ref{sect:mapping}. Note that if an atmospheric temperature is lower than an ambient temperature by 10~K, MiSTI may underestimate the opacity by $\sim$10\% for a opacity of $\sim$0.4, which is approximately a median optical depth of the atmosphere in Atacama at 183$\pm$4.5~GHz.

\subsection{Control System}
 The hardware for control and measurement (e.g., motor drivers, power meter, digital multimeter) was put inside the mount box of MiSTI as mentioned above, whereas a dedicated personal computer (PC) is placed in the container right below the telescope. 
The control software is installed on the PC, and it controls the azimuth and elevation motors, the chopping mirrors, the power meter for IF power measurements, and a digital multimeter for temperature measurements via a local area network and a GPIB interface (GPIB-ENET/100, National Instruments Inc.). See also figure~\ref{system}.
The control software runs on the Linux operating system, and is written in C and Perl languages. Observations are performed by running a script that has azimuth and elevation coordinates for each integration and is prepared in advance. The software reads the script and sends to the motor drivers the number of pulses that corresponds to the pointing coordinates. When the telescope points to the intended direction of the sky, the software starts to acquire and save the output from the power meter. Simultaneously the software measures and records the load temperatures. Currently, observing scripts for standard scans (e.g., skydips, raster-scans and all-sky maps) and the associated reduction software are available.


\section{The Sky Opacity Monitor}
\label{sect:monitor}

MiSTI is operating as the ASTE Sky Monitor, which is open to the public through a web site\footnote{\texttt{http://aste-www.mtk.nao.ac.jp/\symbol{"7E}misti/opacity.html}}. The monitor provides useful information for constructing (sub)mm observing strategies at the Pampa la Bola and Chajnantor sites. The ASTE Sky Monitor includes two monitors: (1) the all-sky opacity monitor, which shows the 183-GHz opacity distribution over a whole sky and (2) the atmospheric window monitor, which offers transmission curves of the atmospheric windows. Here we provide description of those monitors.

\subsection{All-Sky Opacity Mapping}
\label{sect:mapping}

The radiometer measures the power from the sky at each azimuth and elevation, $P_\mathrm{sky}(Az,\,El)$. The integration time at each $Az$ and $El$ is 0.4~sec, which provides a receiver noise level of $\sim$0.05~K (rms). For intensity calibration, we employ the one-load chopper wheel method at the moment and will update it to the dual-load method. The radiometer measures the power from a load, $P_\mathrm{cal}(t)$, once every 50 integrations corresponding to a time duration of $\sim$30~sec. The load is set inside the receiver cabin and its temperature is not actively controlled but monitored. The all-sky mapping typically takes $\sim$40 min, and the monitor is updated hourly. 

When the absorbing layer of the sky is approximately expressed as a plane parallel slab with a finite thickness in height and the atmosphere and load temperatures can be regarded to be the same, the relationship among $P_\mathrm{sky}$, $P_\mathrm{cal}$, $Az$ and $El$ can be described as
\begin{eqnarray}\label{eq:caltau}
\ln{\left[ 1-\frac{P_\mathrm{sky} (Az, El; t)}{P_\mathrm{cal} (t) } \right] } = -\tau_0 \sec{Z} + b, 
\end{eqnarray}
where $Z$ is the zenith angle defined as $(90-El)$~deg, and $\tau_0$ and $b$ are positive constants. The constant $\tau_0$ is called zenith opacity, which is the optical depth of the sky absorbing layer at the zenith ($El =90\arcdeg$). We derive the constants $\tau_0$ and $b$ by fitting the data to a linear function of $\sec{Z}$. The opacity in an arbitrary direction $(Az,El)$ is then described as
\begin{eqnarray}\label{eq:opacitymap}
\tau (Az, El) = b - \ln{\left[ 1-\frac{P_\mathrm{sky} (Az, El; t)}{P_\mathrm{cal} (t) } \right] }.
\end{eqnarray}
We define the opacity fluctuation $\delta\tau(Az,El)$ as
\begin{eqnarray}\label{eq:fluctuation}
   \delta \tau (Az, El) = \frac{\tau(Az, El) - \tau_0 \sec{Z}}{\tau_0 \sec{Z}}.
\end{eqnarray}

\begin{figure*}[t]
\begin{center}
	\FigureFile(130mm,1mm){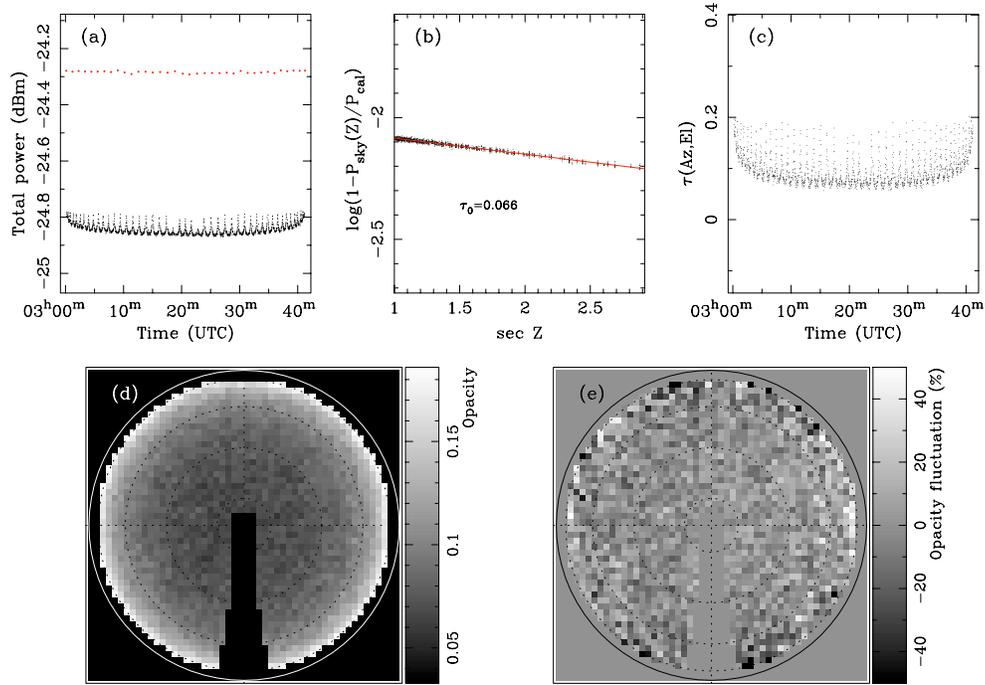}
	\end{center}
		\caption{ An example of the all-sky opacity image taken with MiSTI on 2010 July 1. 
			(a) The total power which the receiver measured on the sky (black dots) and on the 
			room load (red dots), as a function of time in UTC. 
			(b) The fit. The best-fit linear function is shown in the red line. The zenith opacity at 
			183~GHz is $\tau _0^{183} = 0.066$, which is almost the best conditions at Pampa 
			la Bola.
			(c) Calibrated time-stream.
			(d) All-sky opacity map at 183~GHz in the fish-eye lens view defined as 
			Equation~\ref{eq:opacitymap}. The sky hemisphere is orthographically projected to 
			the plane of the paper. The north is top and the east is left. The solid and dotted circles 
			indicate the horizon ($El = 0\arcdeg$), $El = 20, 40, 60$, and $80\arcdeg$. 
			(e) The fluctuations of the all-sky opacity map, $\delta \tau (Az,El)$, which is define as 
			Equation~\ref{eq:fluctuation}. In this measurement, the opacity typically fluctuates 
			by 8\% (rms) at $El > 60\arcdeg$. See the text for details. 
		}\label{diagnostics}
\end{figure*}

\begin{figure*}
	\begin{center}
		\FigureFile(70mm,1mm){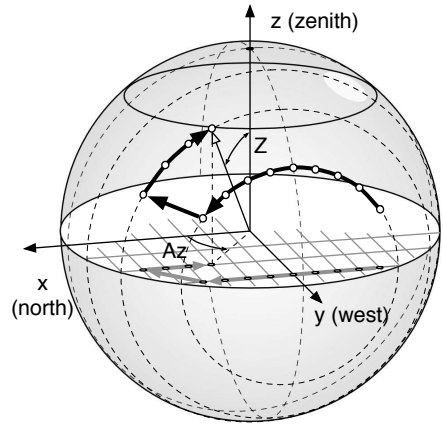}
	\end{center}
	\caption{
		A schematic picture of the scanning pathway of the MiSTI's pointing used in the all-sky opacity 
		measurement. The sphere represents the sky in the horizontal coordinate system, and 
		the white disk perpendicular to the $z$ axis represents the horizon. The open circles on 
		the sphere indicate the direction which MiSTI points to. The positions on the sky hemisphere 
		are selected, such that the corresponding positions orthographically projected to the 
		horizon plane $(x, y)$ are on regular grid points $(x_i, y_j)$, where $i$ and $j$ are integer. 
		The azimuthal and zenith angles are then given by $Az = \tan^{-1}{(y_j/x_i)}$ and 
		$Z = \sin^{-1}{\sqrt{x_i^2+y_j^2}}$. The thick and thin arrows connecting the open circles 
		show the scanning path on the sky and on the horizontal plane (or an image shown in 
		figure~\ref{diagnostics}d).
	}\label{hemisphere}
\end{figure*}

An example of the all-sky measurements and results is presented in figure~\ref{diagnostics}. The scanning pathway is shown in figure~\ref{hemisphere}.

\subsection{Atmospheric Modeling}

The 183~GHz opacity ($\tau_0^{183}$) is a good measure of PWV when the measured sky opacity $\lesssim 1$ as discussed in \S~\ref{sect:waterVapor}. MiSTI exploits the advantage of the 183~GHz spectrum, and provides not only PWV but also the transmission curves of the atmospheric windows by introducing an atmospheric transmission model, \textsc{am} \citep{Paine04}. \textsc{am} is a program for producing model atmospheric emission, absorption, and phase delay at frequencies between a few GHz and a few THz. The model that we use here is optimized for the Chajnantor site (\textsc{am} configuration file, \verb|Chajnantor_lowres.amc|). The model does not consider liquid phase of water. 
The dominant parameters that constrain the atmospheric spectra (or opacity at an arbitrary frequency) are PWV and ground temperature, $T_\mathrm{gr}$, which allows us to estimate PWV from measurements of $\tau _0^{183}$ and $T_\mathrm{gr}$. We use the atmospheric model to calculate $\tau _0^{183}$ for each PWV and $T_\mathrm{gr}$, and then we search for the PWV that accounts for the measured $\tau_0$ and $T_\mathrm{gr}$ using the least-$\chi ^2$ method. 
Once PWV is determined, we can then estimate the opacity spectrum over the mm/submm bands using \textsc{am}.

\subsection{Comparing Measurements at 220 GHz}
\label{sect:comparison}

To assess how correctly the opacity measurement and modeling (i.e., frequency transfer of atmospheric opacity estimates) work, we compared the zenith opacities measured at 183~GHz with those at 220~GHz. The 220~GHz data were retrieved from the data archive of the tipper at the ASTE site \citep{Kohno95}, which have been cross-checked by side-by-side experiments involving the 220~GHz tipper and other instruments such as the NRAO 225~GHz tipper \citep{Radford98} and an FTS \citep{Matsuo98, Matsushita99, Matsushita00}. We estimated the 220~GHz opacities from our 183~GHz measurements by modeling the atmospheric transmission, and compared them with the actual 220~GHz measurements. 

\begin{figure*}
	\begin{center}
		\FigureFile(110mm,1mm){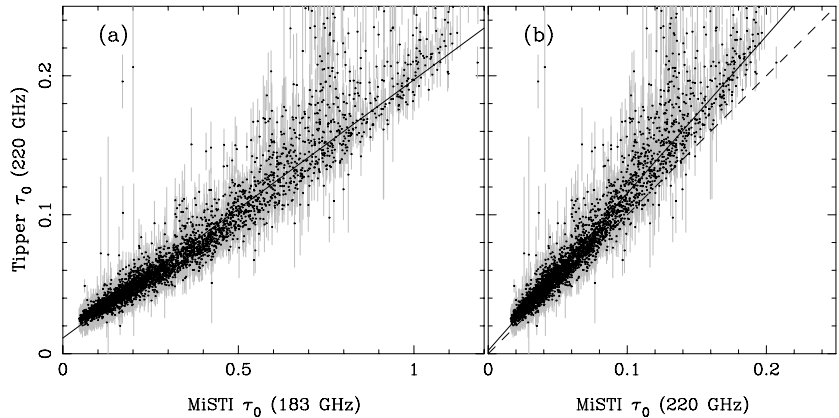}
	\end{center}
		\caption{
			(a) The comparison of zenith opacities measured with the 220~GHz tipper 
			($\tau_0^\mathrm{220}$) and MiSTI ($\tau_0^\mathrm{183}$). The error bars 
			on each plot are calculated from rms of tipper outputs during each of the MiSTI 
			measurements ($\approx$41~min). The solid line shows the best-fit linear function, 
			$\tau_0^\mathrm{220} = 0.186\, \tau_0^\mathrm{183} + 0.011$. 
			(b) The comparison of zenith opacities measured with the 220~GHz tipper 
			($\tau_0^\mathrm{220}$) and those estimated from the MiSTI 183~GHz measurements 
			($\tau_0^\mathrm{220\,MiSTI}$). The solid line shows the best-fit linear function, 
			$\tau_0^\mathrm{220} = 1.132\, \tau_0^\mathrm{220\, MiSTI} + 0.002$. The dashed 
			line indicates $\tau_0^\mathrm{220} = \tau_0^\mathrm{220\, MiSTI}$.
			 See \S~\ref{sect:comparison} for more details.
		}\label{tauplot}
\end{figure*}

Figure~\ref{tauplot}a shows the opacity correlation between 183 and 220~GHz. The data were collected during March to August and October to December in 2008. The correlation is excellent and the correlation coefficient is 0.957. The best-fit linear function is $\tau_0^\mathrm{220} = 0.186\, \tau_0^\mathrm{183} + 0.011$. 
In figure~\ref{tauplot}b, we compare the 220~GHz opacities which are estimated from the 183~GHz measurements, with those actually measured with the tipper. Again, the correlation is very good (the correlation coefficient is 0.958). When the opacities are relatively low ($\tau_0^\mathrm{220} \le 0.08$, or 70\% of the data points in our comparison), the transfered 220~GHz opacities are consistent with the tipper's opacities ($\tau_0^\mathrm{220} = \tau_0^\mathrm{220\, MiSTI}$). This suggests that the 183~GHz measurements of MiSTI are correct, down to the level of any residual systematic errors in the measurements of the 220~GHz tipper.

In contrast, when the opacities are relatively high ($\tau_0^\mathrm{220} \gtrsim 0.1$), the actual 220~GHz measurements clearly exceed the transfered opacities. The best-fit line to all of the data is expressed as $\tau_0^\mathrm{220} = 1.1432\ \tau_0^\mathrm{220\, MiSTI} + 0.002$, suggesting a $\sim$14\% excess of $\tau_0^\mathrm{220}$ to the MiSTI measurements on average. This might be due to a calibration error in load--sky measurements at 183~GHz, which decreases the 183~GHz optical depths, as suggested in Section~\ref{sect:calibration}. However, the calibration scheme, the load material and temperature used in MiSTI are the same as those used in the tipper. It is thus unlikely that the calibration method itself largely affects the discrepancy.

Another more likely origin of the excess can be the effect of water droplets on the optical depth measured with the 220~GHz tipper. It is known that not only water vapor but also liquid water contribute to the optical depth at 220~GHz whereas the 183~GHz opacity depends almost only on pure water vapor (e.g., \cite{Matsushita00}). It is thus likely that the liquid phase of water more affects on $\tau_0^\mathrm{220}$ than $\tau_0^\mathrm{183}$. In the next section, we discuss the effect from water droplets contained in visible clouds.


\section{Discussions}
\label{sect:discussions}

\begin{figure*}
	\begin{center}
		\FigureFile(80mm,1mm){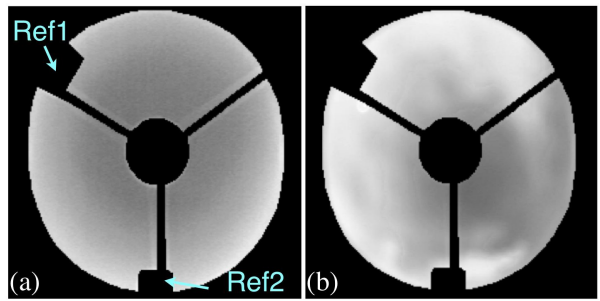}
	\end{center}
	\caption{
		Examples of emissivity distribution derived from the images taken with the IR cloud 
		monitor at the ASTE site. 
		(a) under clear sky condition $(\tau _{\mathrm{IR}} \approx 0.25)$, and  
		(b) under cloudy weather $(\tau _{\mathrm{IR}} \approx 0.5)$. 
		Two aluminum blocks ``Ref1'' and ``Ref2'' are placed at the edge of the main mirror to 
		be used as brightness standards, but only ``Ref1'' was used in the analyses shown in this paper. 
		Note that the shadows by these blocks, sub-reflector and its pillars are masked in these images. 
	}\label{irmon}
\end{figure*}

The sky data obtained with the IR cloud monitor at the ASTE site can be used to examine the effect of the liquid component of water on the correlation between $\tau_0^{\mathrm{183}}$ and $\tau_0^{\mathrm{220}}$. The cloud monitor consists of a gold-plated Cassegrain-type mirror, a commercial mid-IR camera which has $320\times 240$ image pixels, and a Linux based computer for control and data gathering. The cloud monitor can automatically take and store a fish-eye image of the sky every 5 minutes (figure~\ref{irmon}). The camera is sensitive to the wavelength range of 8--12~$\mu$m. Before installing the camera to the ASTE site, we measured its characteristics in the laboratory by taking images of an objects with various temperature ($T_{\mathrm{obj}}$), and we confirmed that the output of each pixel is quite proportional to 
\begin{equation}
\int _{8\mu\mathrm{m}}^{12\mu\mathrm{m}}B_{\lambda}(T_{\mathrm{obj}}) d\lambda
\end{equation}
in the range of $T_{\mathrm{obj}} = 255$--330~K, where $B_{\lambda}(T)$ is the Planck function. In each frame of the sky image, the IR emission from a black-painted aluminum block placed at the edge of the main mirror is also taken simultaneously (see figure~\ref{irmon}). The temperature of the block is always monitored, and its emission is used as the brightness standard when the emissivity distribution of the sky is calculated. The obtained sky emissivity is always above 0; Even when there are no clouds, the emissivity is uniform and its typical value is 0.23--0.33, or 0.25--0.4 in optical depth at these IR wavelengths ($\tau _{\mathrm{IR}})$. This is mainly because there is an emission band of atmospheric ozone in this wavelength range. The measured accuracy is also limited by the accumulation of dirt on the mirror surface. More detailed description about the system can be found in the papers reported by \citet{Takato03}, \citet{Suganuma07}, and \citet{Miyata08}. 

\begin{figure*}
	\begin{center}
		\FigureFile(140mm,1mm){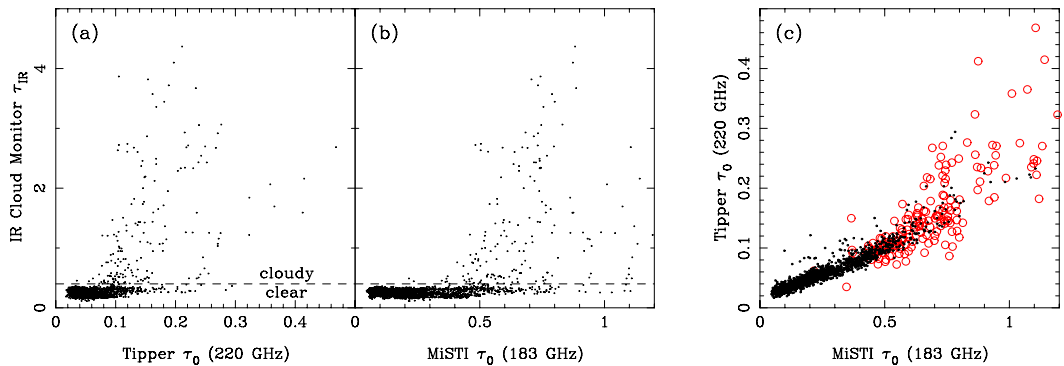}
	\end{center}
	\caption{Comparisons among the data taken with the tipper working at 220~GHz, 
		MiSTI and the IR cloud monitor during the period from April to August in 2008; 
		only data that were taken simultaneously with these three monitors are included.  
		The time resolution of the data  is smoothed to be 1 hour. 
		(a) Comparisons between the 220~GHz optical depth toward zenith  ($\tau_0^{\mathrm{220}}$) 
		with the tipper and  the IR optical depth averaged over the regions where the elevation is 
		greater than $60^{\circ} ~(\tau_{\mathrm{IR}})$ derived from the images with the IR 
		cloud monitor. There are 2279 data points in total. The dashed line indicates the border 
		between clear and cloudy conditions ($\tau_{\mathrm{IR}} = 0.4$).
		(b) Comparisons between the 183~GHz optical depth toward zenith  ($\tau_0^{\mathrm{183}}$) 
		with MiSTI and  the $\tau_{\mathrm{IR}}$. The number of the data points is the same as (a). 
		(c) Comparisons between $\tau_0^{\mathrm{183}}$  and $\tau_0^{\mathrm{220}}$ 
		when the sky is clear ($\tau_{\mathrm{IR}} \leq 0.4$, black dots) and cloudy ($\tau_{\mathrm{IR}} > 0.4$, 
		red open circles). There are 2108 and 171 data points for the $\tau_{\mathrm{IR}} \leq 0.4$ 
		and $\tau_{\mathrm{IR}} > 0.4$ conditions, respectively. 
		}\label{compirmm}
\end{figure*}

Figures~\ref{compirmm}a and  \ref{compirmm}b show the comparison of optical depths at mm wavelengths and that at mid-IR during the period from April to August in 2008. These figures contain data that were taken only when all three environmental monitors at the ASTE site (i.e., MiSTI, tipper, and IR cloud monitor) were working. The time resolution of the data in figure~\ref{compirmm} is smoothed to be 1 hour, which corresponds to 1.5 times the time resolution of MiSTI. When $\tau_0^{\mathrm{220}} < 0.05$ or $\tau_0^{\mathrm{183}} < 0.4$, $\tau_{\mathrm{IR}} = 0.25$--0.4, indicating that the sky was clear, while when  the optical depths at mm wavelengths are large ($\tau_0^{\mathrm{220}} > 0.05$ or $\tau_0^{\mathrm{183}} > 0.4$), $\tau_{\mathrm{IR}}$ is large but also shows large scatter (figures~\ref{compirmm}a and \ref{compirmm}b). $\tau_0^{\mathrm{183}}$ changes by an order of magnitude even under clear sky conditions, suggesting that the observed frequency used by MiSTI can give us critical information to select the best conditions for submm observations. 
Figure~\ref{compirmm}c presents comparisons between  $\tau_0^{\mathrm{220}}$ and $\tau_0^{\mathrm{183}}$ under clear sky (i.e. $\tau_{\mathrm{IR}} \leq 0.4$, dots) and cloudy conditions ($\tau_{\mathrm{IR}} > 0.4$, open circles). These comparisons reveal the followings: (i) $\tau_0^{\mathrm{220}}$ is always greater than 0.05 when the sky is cloudy, (ii) the correlation is much tighter under clear sky conditions, and (iii) $\tau_0^{\mathrm{220}}$ under cloudy sky shows large scatter and is even larger than the value expected from the correlation under clear sky. These results imply that the liquid component of water affects $\tau_0^{\mathrm{220}}$ more than $\tau_0^{\mathrm{183}}$, as discussed by \citet{Matsushita00}. The correlation under clear sky shown in figure~\ref{compirmm}c shows  a bend at ($\tau_0^{\mathrm{183}}, \tau_0^{\mathrm{220}}$) = (0.4, 0.06) and the slope gets steeper in the regions of larger $\tau_0^{\mathrm{183}}$, which is also seen in figure~\ref{tauplot}a. This may also be explained by more significant contribution of liquid water to $\tau_0^{\mathrm{220}}$. 


\section{Summary}
\label{sect:summary}

We have developed a small mm-wave telescope equipped with a 25-cm diameter dish, named the Millimeter Sky Transparency Imager (MiSTI). It measures emission from atmospheric water vapor at frequencies of $183.3 \pm 4.5$~GHz, which maximize the sensitivity to tiny changes in the PWV content of the atmosphere. MiSTI observes the water vapor distribution at a spatial resolution of $0.5 \arcdeg$, and it is sensitive enough to detect an excess path length of $\lesssim 0.05$~mm for an integration time of 1~s.
By comparing the measurements of MiSTI at 183~GHz with those from the 220~GHz tipper, we validate that the 183~GHz measurements of MiSTI are correct, down to the level of any residual systematic errors in the measurements at 220~GHz. When the mm opacities are large ($\tau_0^\mathrm{183} > 0.4$), optical depths measured at 220~GHz start to exceed those at 183~GHz. Given the comparison between the mm and IR opacities, this excess is likely attributed to water droplets in the sky, which contribute more largely to 220~GHz opacity than 183~GHz. 
Currently, MiSTI provides real-time  (every 1~hr) monitoring of the all-sky opacity distribution and atmospheric transmission curves in the (sub)mm through the internet, allowing us to know the (sub)mm sky conditions at Atacama. 
It is important to investigate $\sim$10--100~m scale structure of water vapor distribution because the small-scale fluctuations of the water vapor affect the natural seeing of submm single-dish telescopes and interferometers. We plan to update the functionality of MiSTI to further increase its sensitivity and measurement efficiency of such fluctuations.

\bigskip
We are grateful to the referee for fruitful comments. 
We thank M.\ Uehara, T.\ Okuda, N.\ Mizuno, and the ASTE team for the support in installation and operation of MiSTI.
We also thank W.\ Kimura, T.\ Miyata and H.\ Matsuo for contributions to the development of the IR cloud monitor.
YT thank A.\ Endo and M.\ Kamikura for fruitful discussions. 
We acknowledge to the Advanced Technology Center (ATC) at NAOJ for allowing the use of the facilities.
This work was financially supported by
an ATC/NAOJ program for joint research and development and 
Grant-in-Aid for Scientific Research (A) (No.\ 18204017). 
YT was financially supported by the Japan Society for the Promotion of Science (JSPS) for Young Scientists.


\clearpage

\begin{table*}
	\caption{The specifications of MiSTI.}\label{tab:specification}
	\begin{center}
	\begin{tabular}{lll}
	\hline \hline
	Parameter & & Value\\
	\hline
	\multicolumn{3}{c}{Main Reflector}\\
	\hline
	Optics & & offset paraboloid \\
	Diameter & $D$ & 250~mm \\
	Confocal distance & & $\sim$30~m \\
	Beam size (HPBW) & $\Delta \theta$ & 30~arcmin \\
	Surface accuracy & & $< 20~\micron$ \\
	Edge taper & $T_\mathrm{e}$  & $-$20~dB \\
	Predicted spillover efficiency & & 0.99 \\
	\hline
	\multicolumn{3}{c}{Receiver}\\
	\hline
	Receiver system & & heterodyne radiometer \\
	Observing frequency	& $\nu_\mathrm{obs}$ & $183.3 \pm 4.5$~GHz \\
	Intermediate frequency & $\nu_\mathrm{IF}$ 	& 4.0--5.0~GHz \\
	DSB band width & $B$ & 2.0 GHz \\
	Receiver noise temperature & $T_\mathrm{RX}$(DSB)	& $\approx$900 K \\
	Gain stability & $\Delta G/G$ & $\lesssim 10^{-4}$ \\
	\hline
	\multicolumn{3}{c}{Misc.}\\
	\hline
	Calibration method & & one/dual-load chopper wheel method \\
	Mount & & alt-azimuth mount \\
	Rotation speed & & 10~deg~s$^{-1}$ \\
	Beam switch & & wobbling main reflector (1~Hz, $30'$)\\
	\hline
	\end{tabular}
	\end{center}
\end{table*}


\begin{thebibliography}{}
	\bibitem[Altenhoff et al.(1987)]{Altenhoff87} Altenhoff, W.~J., Baars, J.~W.~M., Wink, J.~E., \& Downes, D.\ 1987, \aap, 184, 381
	\bibitem[Beaupuits et al.(2005)]{Beaupuits05} Beaupuits, J.~P.~P., Rivera, R.~C., \& Nyman, L.\ 2005, ALMA Memo, 542
	\bibitem[Carilli \& Holdaway(1999)]{Carilli99} Carilli, C.~L., \& Holdaway, M.~A.\ 1999, Radio Science, 34, 817 
	\bibitem[Delgado et al.(2001)]{Delgado01} Delgado, G., Nyman, L., Ot\'{a}rola, A., Hills, R., \& Robson, Y.\ 2001, ALMA Memo, 361
	\bibitem[Ezawa et al.(2004)]{Ezawa04} Ezawa, H., Kawabe, R., Kohno, K., \& Yamamoto, S.\ 2004, \procspie, 5489, 763 
	\bibitem[Ezawa et al.(2008)]{Ezawa08} Ezawa, H., et al.\ 2008, \procspie, 7012, 6
	\bibitem[Kohno et al.(1995)]{Kohno95} Kono, K., Kawabe, R., Ishiguro, M., Kato, T., \& Bronfman, L.\ 1995, LMSA Technical Memo, 1
	\bibitem[Matsuo et al.(1998)]{Matsuo98} Matsuo, H., Sakamoto, A., \& Matsushita, S.\ 1998, \pasj, 50, 359
	\bibitem[Matsushita et al.(1999)]{Matsushita99} Matsushita, S., Matsuo, H., Pardo, J.~R., \& Radford, S.~J.~E.\ 1999, \pasj, 51, 603 
	\bibitem[Matsushita et al.(2000)]{Matsushita00} Matsushita, S., Matsuo, H., Sakamoto, A., \& Pardo, J.~R.\ 2000, \procspie, 4015, 378 
	\bibitem[Miyata et al.(2008)]{Miyata08} Miyata et al.\ 2008, \procspie, 7012,  140
	\bibitem[Paine(2000)]{Paine00} Paine, S., Blundell, R., Papa, D.~C., Barrett, J.~W., \& Radford, S.~J.~E.\ 2000, \pasp, 112, 108 
	\bibitem[Paine(2004)]{Paine04} Paine, S.\ 2004, SMA Technical Memo, 152
	\bibitem[Pardo et al.(2001)]{Pardo01} Pardo, J.~R., Cernicharo, J., \& Serabyn, E.\ 2001, IEEE Proceedings, 49, 1683
	\bibitem[Radford \& Holdaway(1998)]{Radford98} Radford, S.~J., \& Holdaway, M.~A.\ 1998, \procspie, 3357, 486 
	\bibitem[Robson et al.(2002)]{Robson02} Robson, Y., Hills, R., Richer, J., Delgado, G., Nyman, L., Ot{\'a}rola, A., \& Radford, S.\ 2002, Astronomical Site Evaluation in the Visible and Radio Range, 266, 268 
	\bibitem[Sebag et al.(2008)]{Sebag08} Sebag, J., Krabbendam, V.~L., Claver, C.~F., Andrew, J., Barr, J.~D., \& Klebe, D.\ 2008, \procspie, 7012,  135
	\bibitem[Shamir et al.(2005)]{Shamir05} Shamir, L., \& Nemiroff, R.~J.\ 2005, \pasp, 117, 972 
	\bibitem[Suganuma et al.(2007)]{Suganuma07} Suganuma, M., et al.\ 2007, \pasp, 119, 567
	\bibitem[Takato et al.(2003)]{Takato03} Takato, N., Okada, N., Kosugi, G., Suganuma, M., Miyashita, A., \& Uraguchi, F.\ 2003, \procspie, 4837, 872 
	\bibitem[Thompson, Moran, \& Swenson(2001)]{Thompson01} Thompson, A.~R., Moran, J.~M., \& Senson, Jr., G.~W.\ 2001, Interferometry and Synthesis in Radio Astronomy (2nd ed.; New York, NY: John Wiley \& Sons, Inc.)
	\bibitem[Wiedner et al.(2001)]{Wiedner01} Wiedner, M.~C., Hills, R.~E., Carlstrom, J.~E., \& Lay, O.~P.\ 2001, \apj, 553, 1036 
	\bibitem[Wootten \& Thompson(2009)]{Wootten09} Wootten, A., \& Thompson, A.~R.\ 2009, IEEE Proceedings, 97, 1463 
\end{thebibliography}
\end{document}